\documentclass[aps,nofootinbib,noshowpacs,showkeys,preprintnumbers,amsmath,amssymb]{revtex4}
\usepackage{graphicx,color}
\usepackage{graphics}
\usepackage{rotating}
\usepackage{amssymb}

\usepackage{color}
\usepackage{xcolor}
\usepackage{epsfig}
\usepackage{dcolumn}% Align table columns on decimal point
\usepackage{bm}% bold math
\usepackage{multirow}

\def\be{\begin{equation}}
\def\ee{\end{equation}}
\def\ba{\begin{eqnarray}}
\def\ea{\end{eqnarray}}
\def\bea{\begin{eqnarray}}
\def\eea{\end{eqnarray}}
\def\bes{\begin{subequations}}
\def\ees{\end{subequations}}
\def\bear{\begin{array}}
\def\eear{\end{array}}

\newcommand{\A}{{\mathcal{A}}}
\newcommand{\tA}{{\widetilde {\mathcal{A}}}}

\newcommand{\mh}{{\mathfrak h}}
\newcommand{\ma}{{\mathfrak a}}
\newcommand{\mH}{{\mathfrak H}}
\newcommand{\mA}{{\mathfrak A}}

\newcommand{\tmA}{{\widetilde  {\mathfrak A}}}

\newcommand{\tK}{{\widetilde K}}

\newcommand{\tB}{{\widetilde B}}

\newcommand{\cO}{{\cal O}}
\newcommand{\cD}{{\cal D}}

\newcommand{\cDo}{{{\cal D}^{(1)}}}
\newcommand{\cF}{{\cal F}}
\newcommand{\tcF}{{\widetilde {\cal F}}}

\newcommand{\tcK}{{\widetilde {\cal K}}}

\newcommand{\MSbar}{\overline{\rm MS}}  
  
\newcommand{\bam}{\overline{m}}  

\begin{document}
%\preprint{USM-TH-xxx}

\title{Renormalon-based resummation for B(D) mesons}
\author{C\'esar Ayala$^a$}
\email{c.ayala86@gmail.com}
\author{Gorazd Cveti\v{c}$^b$}
\email{gorazd.cvetic@gmail.com}
%\author{Reinhart K\"ogerler$^c$}
%\email{koeg@physik.uni-bielefeld.de}
\affiliation{$^a$Departamento de Ingenier\'ia y Tecnolog\'ias, Sede La Tirana, Universidad de Tarapac\'a, Av.~La Tirana 4802, Iquique, Chile}
\affiliation{$^b$Department of Physics, Universidad T{\'e}cnica Federico Santa Mar{\'\i}a, Avenida España 1680, Valpara{\'\i}so, Chile} 
%\affiliation{$^c$Department of Physics, Universit\"at Bielefeld, 33501 Bielefeld, Germany}

\date{\today}

\begin{abstract}
We apply a previously developed method of renormalon-based resummation of spacelike and timelike QCD observables, to evaluate of the values of the pole masses $m_q$ of $q=b$ and $c$ quarks, using as input the knowledge of the values of the corresponding ${\overline{\rm MS}}$ masses ${\overline m}_q \equiv {\overline m}_q({\overline m}_q^2)$. The evaluation also uses the knowledge of the first few coefficients of the perturbation expansion of $m_q$ (i.e., of $m_q/{\overline m}_q$), as well as the known renormalon structure of that expansion. In the evaluation, we use the timelike QCD running coupling based on a specific holomorphic spacelike QCD coupling, in order to avoid additional ambiguities due to the Landau poles of the usual perturbative coupling. The principal IR regulator parameter of the coupling is varied in an expected range. We also reevaluate the chromomagnetic Wilson coefficient of heavy quark ${\hat C}(m_q^2)$, and extract values of several corresponding hadronic parameters.
\end{abstract}

\maketitle

\section{Introduction}
\label{sec:intr}

The description and evaluatation of QCD phenomenology at moderately low energies $Q \lesssim 1$ GeV remains a longstanding challenge, because the perturbation approach that is working fine for higher energies starts failing once we approach, from above, the range $Q \sim 1$ GeV. This is connected with the factorial divergence of the coefficients at large orders $n$ in the perturbation expansion of QCD observables ($d_n \sim n!$, also referred as renormalon growth \cite{Renormalons}), and with the Landau singularities of the running perturbative couplings $a(Q^2) \equiv \alpha_s(Q^2)/\pi$ at low spacelike momenta $0 \leq Q^2 \lesssim 1 \ {\rm GeV}^2$ (where: $Q^2 \equiv -q^2=-(q^0)^2 + {\vec q}^2$).

Landau singularities in the perturbative QCD (pQCD) running coupling $a(Q^2)$ are a consequence of the use of perturbative $\beta(a)$ function. The problem of Landau singularities is addressed by replacing the (pQCD-)coupling $a(Q^2)$ by an (AQCD-)coupling $\A(Q^2)$ that has no Landau singularities, i.e., it is a holomorphic (analytic) function in the complex $Q^2$-plane that excludes most of the negative (timelike) axis, $Q^2 \in \mathbb{C} \backslash (-\infty,-M^2_{\rm thr})$, where $M_{\rm thr}$ ($ \sim m_{\pi}$) is a threshold scale.\footnote{
For some works on the QCD running couplings without Landau singularities, we refer to \cite{Shirkov,SMS,ShirRev,FAPT,SMnu,BakRev} for the Minimal Analytic couplings (MA; named also (F)APT); \cite{2dQCD1,2dQCD2,3dAQCD} for 2$\delta$ and 3$\delta$AQCD coupling; and for various applications, cf.~\cite{ShirkEPJC,Nest1,Nest2,NestBook,GCAK,ACKS,KotBSR,CAGCUps,CASM,Mirj,Nestamu,amu3dAQCD,GGKT,Kotikov}.} 
Certainly, for consistency, at $|Q^2| > 1 \ {\rm GeV}^2$, $\A(Q^2)$ should practically coincide with the underlying $a(Q^2)$. The coupling $\A(Q^2)$ qualitatively reflects the holomorphic properties of spacelike QCD observables $\cD(Q^2)$ (such as correlators, structure functions, etc.) as dictated by the general principles of Quantum Field Theories \cite{BogShir,Oehme}. On the other hand, most of the QCD observables are timelike, $\cF(\sigma)$ (where $\sigma >0$), such as cross-sections, decay widths, and quantities associated with hadronic masses. To evaluate them consistently, we should then use the timelike coupling $\mA(\sigma)$ associated with the mentioned holomorphic spacelike coupling $\A(Q^2)$.

In the mentioned framework with the holomorphic (i.e., free of Landau singularities) spacelike coupling $\A(Q^2)$, the problem of the renormalon growth ($d_n \sim n!$) and the associated divergence of the perturbation expansion, was addressed in a specific renormalon-motivated resummation approach \cite{renmod},\footnote{This approach is an extension of the resummation \cite{Neubert} from the one-loop level to any loop level.}
where the spacelike observable $\cD(Q^2) = \sum_{n \geq 0} d_n a(Q^2)^{n+1}$ was evaluated as an integral of a characteristic function $F_{\cD}(t)$ and the running coupling $a(t Q^2)$ [or, in AQCD: $\A(t Q^2)$], where the characteristic function was determined by the renormalon structure of $\cD(Q^2)$ and by the first few (exactly known) expansion coefficients $d_n$. Applications of this method, using either $a(t Q^2)$ or $\A(t Q^2)$, were made for evaluation of the Adler function \cite{renmod,amu3dAQCD} and Bjorken polarised sum rule \cite{PLB848,NPB1007}. This resummation method was then extended in \cite{JPG2026} to the case when the perturbation expansion of $\cD(Q^2)$ is not necessarily in integer powers, $\cD(Q^2) = \sum_{n \geq 0} d_n a(Q^2)^{n+\nu_0}$ (i.e., $\nu_0 \not=1$, noninteger), and to the evaluation the (associated) timelike observables $\cF(\sigma)$.\footnote{
We note that another method of resummation for spacelike and timelike observables was constructed in \cite{SM1,SMnu,SMnuaux}. We refer to \cite{JPG2026} for a short discussion of this method.}
In \cite{JPG2026}, as an illustration of the method, we applied it to the evaluation of the chromomagnetic Wilson coefficient ${\hat C}(\sigma)$ of heavy quark ($\sigma=m_q^2$), which is a timelike renormalisation invariant QCD quantity with $\nu_0 \not=1$ (noninteger).

In this work, we apply this method in Sec.~\ref{sec:mq} to the evaluation of the pole mass $m_q$ ($q=c,b$), which we regard as a timelike quantity $m_q(\sigma)$ at $\sigma = \bam_q^2$, where $\bam_q \equiv \bam_q(\bam_q^2)$ are the (known) values of the $\MSbar$ mass of quark $q$. In Sec.~\ref{sec:HQETdet} we then reevaluate the chromomagnetic Wilson coefficient ${\hat C}(\sigma)$, at the obtained pole mass values $\sigma=m_c^2, m_b^2$. \textcolor{black}{In both the evaluation of $m_b$ and ${\hat C}(m_b^2)$, we use for the number of effective quark flavours $n_f$ the value $n_f=3$ [as we also use for $m_c$ and ${\hat C}(m_c^2)$], in order to achieve decoupling of the charm mass effects.\footnote{In \cite{JPG2026} we used $n_f=4$ for ${\hat C}(m_b^2)$.} This seems to be a typical requirement for quantities that are dominated by the IR renormalon $u=1/2$.}
Further, in that Sec.~\ref{sec:HQETdet} we also extract values of various associated hadronic mass parameters: $\Lambda_{\rm eff}$, ${\hat{\mu}}_G^2$, $\mu_{\pi}^2$ and $\bar \Lambda$. In Sec.~\ref{sec:concl} we then compare the obtained results with those of other works and make a summary of conclusions.

\section{Pole mass from $\overline{\rm MS}$ mass}
\label{sec:mq}

Let $\bam_q(\mu^2)$ denote the $\overline{\rm MS}$ running mass of the quark $q$ ($q=c,b$). A common choice is the scale-invariant point $\mu=\bam_q(\bam_q^2)$, abbreviated as $\bam_q \equiv \bam_q(\bam_q^2)$.

The relation for the pole mass $m_q$, scaled by the value of the mass $\bam_q$, can be written as a perturbation series
\begin{equation}
{\cal F}_q(\sigma) \equiv \frac{3}{4} \left( \frac{m_q(\sigma)}{\bam_q} - 1 \right) = {\bar a}(\sigma) + \sum_{n=1}^{\infty} ({\bar f}_q)_n {\bar a}(\sigma)^{n+1},
\label{Fqexp}
\end{equation}
where $a(\mu^2) \equiv \alpha_s(\mu^2)/\pi$, and the bars over the coupling and the coefficients indicate that the series is in the $\MSbar$ scheme.
We introduced in the quantity $m_q$ a dependence on the timelike squared energy $\sigma$, and the usual pole mass\footnote{\textcolor{black}{In \cite{KatEPJC}, a somewhat different approach is taken, where the $\MSbar$ mass is also running, ${\bam_q}(\sigma)$. Stated otherwise, there the running pole mass is $M_q(\sigma) =  {\bam_q}(\sigma) [1 + (4/3) {\cal F}_q(\sigma)]$, while in our case it is $m_q(\sigma) = {\bam_q} [1 + (4/3) {\cal F}_q(\sigma)]$ where ${\bam_q}={\bam_q}(\bam_q^2)$. In both cases, the pole mass is $M_q(\bam_q^2)=m_q(\bam_q^2)$.}}
is then at $\sigma=\bam_q^2$
\be
m_q \equiv m_q(\sigma=\bam_q^2).
\label{mqpole} \ee
We can use the expansion (\ref{Fqexp}) in another scheme\footnote{
The change of scheme, from the $\MSbar$ scheme, is defined, in the convention used here, as the change coming solely from the change of the $\beta_j$ coefficients (${\bar \beta}_j \mapsto \beta_j$; $j \geq 2$) of the usual perturbative beta-function, $d a/d \ln \mu^2 = - \sum_{j \geq 0} \beta_j a^{j+2}$. For $n_f=3$ we have $\beta_0=9/4$ and $\beta_1=4$.}
and at another renormalisation scale $\mu^2=\kappa \sigma$, and in that case the expansion coefficients $({\bar f}_q)_n$ correspondingly change to $(f_q)_n(\kappa)$
\begin{equation}
{\cal F}_q(\sigma) \equiv \frac{3}{4} \left( \frac{m_q(\sigma)}{\bam_q} - 1 \right) = a(\kappa \sigma) + \sum_{n=1}^{\infty} (f_q)_n(\kappa) \; a(\kappa \sigma)^{n+1}.
\label{Fqexpgen}
\end{equation}
Here, $a(\kappa \sigma) = a(\mu^2=\kappa \sigma; \beta_2, \beta_3,\ldots)$ is the coupling in the considered scheme ($\beta_2,\beta_3,\ldots$) and at the considered scale $\mu^2$. 
The series is known up to $\mathcal{O}(a^4)$ in the massless-light-quark limit: 
\cite{f0,f1,f2} for $({\bar f}_q)_n$, $n=0,1,2$, respectively and \cite{f3Smi1,f3Smi2,f3Zum} for ${\bar f}_3$.

In heavy-flavour applications, the finite-mass effects (e.g.\ charm loops in the bottom pole mass $m_b$) are known to at least $\mathcal{O}(a^3)$, as discussed in \cite{ACPmass}: see Refs.~\cite{f1} and \cite{dmc2ex} for $\cO(a^2)$ and $\cO(a^3)$ corrections, respectively. The IR behaviour of the $\cO(a^2)$ charm loops correction was discussed in \cite{dmc1IR}, and a linear approximation of the $\cO(a^3)$ charm loops correction was obtained in \cite{dmc2lin}. 

When fixing the renormalization scale to $\mu^2=\bam_q^2$ (i.e., $\kappa=1$), the known $({\bar f}_q)_n$ coefficients in the $\MSbar$ scheme in the polynomial form in $n_f$ are as follows: 
\bes
\label{barfj}
\bea
  ({\bar f}_q)_1(n_f) &=& 10.0823 - 0.781 \; n_f,
\label{bf1}\\
 ({\bar f}_q)_2(n_f) &=& 142.792 - 19.9913 \; n_f + 0.489521 \; n_f^2,
\label{bf2}\\
  ({\bar f}_q)_3(n_f) & \approx & 2671.65 - 558.967 \; n_f + 32.5472 \; n_f^2 - 0.508606 \; n_f^3.
\label{bf3}
\eea \ees      
Here, $n_f$ is the number of active (light) flavors. \textcolor{black}{In \cite{KatEPJC,KatJETP}, higher order coefficients have been estimated.}
For $m_c$ [i.e., $\cF_c(\bam_c^2)$] we set $n_f=3$. For $m_b$ [i.e., $\cF_b(\bam_b^2)$] we also set $n_f=3$. \textcolor{black}{This is so for the following reason (as explained in \cite{ACPmass}): in the mass $m_b = m_b^{(n_f=4)} + \delta^{(0)}_{m_c}$, the correction $\delta^{(0)}_{m_c}$ due to $m_c \not= 0$ represents a large divergent series; however, when we reformulate this same sum as $m_b^{(n_f=3)} + \delta^{(\infty)}_{m_c}$, where $\delta^{(\infty)}_{m_c}$ is the correction due to $m_c \not= \infty$, the latter correction turns out to be very small, practically zero.}\footnote{\label{BBB}
This phenomenon appears to hold in general for the bottom quark mass-related quantities with IR renormalon $u=1/2$, because, as argued in \cite{BBB}, the natural scale in the loop integrals is not $m_b$ but $m_b e^{-(n+1)}$, where $(n+1)$ is the loop order [in our notation, cf.~Eq.~(\ref{Fqexp})], and this scale is in general well below $m_c$.}  

For $n_f=3$, we have the value ${\bar f}_3=1273.9$, which is the value converted from the central value of \cite{f3Smi2} (${\bar f}_3=1277.0$, with $n_f=3$ and $n_h=1$) to the theory with the coupling with $n_f=3$ massless quarks only, according the approach applied in \cite{f3Zum} [cf.~Eq.~(3) there]. 

The evaluation of the perturbation series Eq.~(\ref{Fqexpgen}) is affected by the $u=1/2$ IR renormalon that makes the pole mass $m_q$ ambiguous at $\mathcal{O}(\Lambda_{\rm QCD})$. This is emphasized in the HQET/OPE discussion, and it is why many precision analyses usually rewrite OPEs \textcolor{black}{in terms of a short-distance mass: $\MSbar$ \cite{f0,f1,f2,f3Smi1,f3Smi2,f3Zum,KatEPJC}, kinetic \cite{Bigi:1994ga,Bigi:1996si,Hayashi:2022hjk,HFLAV:2022esi}. PS \cite{Hoang:2000yr}, 1S \cite{Hoang:1998hm,Hoang:1998ng,Bauer:2004ve}, MSR \cite{Hoang:2017suc}, etc.; or use a renormalon-subtracted definition: RS \cite{ACPmass,Pineda:2001zq}, PV \cite{AyaPin2019,AyaPin2026}, FTRS \cite{Hayashi:2021vdq}, DSRS \cite{Hayashi:2023fgl}, etc.}

In this work we consider two dominant renormalons of $\cF_q$, i.e., at $u=1/2$ and $u=-1$. Since the $u=1/2$ IR renormalon is expected to dominate, we include subleading contribution of this renormalon. The Borel structure of ${\cal F}_q$ at those renormalons is known
\begin{equation}
  B[\cF_q](u)=\left[ \frac{\cF_{1/2}^{(0)}}{(1/2-u)^{1+\beta_1/(2\beta_0^2)}}+\frac{\cF^{(1)}_{1/2}}{(1/2-u)^{\beta_1/(2\beta_0^2)}}+\frac{\cF_{-1}}{(1+u)^{1-\beta_1/\beta_0^2}} \right].
 \label{BFq} 
\end{equation}
We point out that the (mass) quantity $\cF_q$ that we want to evaluate, Eq.~(\ref{Fqexpgen}), is a timelike quantity, and its perturbation expansion starts with the term $\sim a^{\nu_0}$ where, in the specific case, $\nu_0=1$. We will now apply the general method for a renormalon-motivated evaluation of this quantity as described in Ref.~\cite{JPG2026}. That method was constructed there for spacelike and timelike quantities with general leading-power index $\nu_0$.\footnote{It was based on the previous work \cite{renmod} that worked generally for spacelike quantities with $\nu_0=1$.} We refer for all the details of the method to that work.

The Borel transform (\ref{BFq}) then implies, according to Theorem 2 of \cite{JPG2026}, that the modified Borel transform $\tB$ of $\cF_q(\sigma)$ has the form\footnote{\textcolor{black}{From the proof of Theorem 2 (in \cite{JPG2026}), it is evident that there the case ${\tilde s}_0 \to 0$ corresponds to: $B[\cF_q](u) \sim 1/(p -u)^{p \beta_1/\beta_0^2}$ and $\tB[\cF_q](u) \sim \ln(1 - p/u)$.}}
\be
\tB[\cF_q](u) = \left[ \frac{\tcF_{1/2}^{(0)}}{(1/2-u)^{1}} + \tcF^{(1)}_{1/2} \ln (1 -2 u) +\frac{\tcF_{-1}}{(1+u)^{1}} \right].
\label{tBFq}
\ee
Then, according to Theorem 5 of Ref.~\cite{JPG2026}, the modified Borel transform $\tB$ of the associated spacelike quantity $\cD_q(Q^2)$ is\footnote{
Since now we have $\nu_0=1$, the quantities $\cD$ and $\cD^{(1)}$ (and: $\cF$ and $\cF^{(1)}$) appearing in Ref.~\cite{JPG2026} coincide with each other: $\cD=\cD^{(1)}$ (and $\cF=\cF^{(1)}$).}
\bes
\label{tBD}
\bea
\tB[\cD_q](u) & = & \frac{\pi u}{\sin (\pi u)} \tB[\cF_q](u)
\label{tBDa} \\
& = &
\left[ \frac{\tcK_{1/2}^{(0)}}{(1/2-u)^{1}} + \tcK^{(1)}_{1/2} \ln (1 -2 u) + \tcK_{-1} \left( \frac{1}{(1+u)^2} - \frac{1}{(1+u)} \right) \right].
\label{tBDb} \eea \ees
In Eq.~(\ref{tBDb}) we used the expansion of the factor $(\pi u)/\sin(\pi u)$ around the point $u=1/2$ for the first two term, and $u=-1$ for the last term
\bes
\label{expfact}
\bea
\frac{\pi u}{\sin (\pi u)} &=& \frac{\pi}{2} + \cO(1/2-u),
\label{expfacta} \\
\frac{\pi u}{\sin (\pi u)} &=& \frac{1}{(1+u)} -1 + \cO(1+u).
\label{expfactb} \eea \ees
Everywhere in the (modified) Borel transforms we neglected the terms with positive powers of $(1/2-u)$ and $(1+u)$ (as they are less singular or nonsingular),
and we regarded that the renormalisation scale parameter $\kappa$ is $\kappa=1$. As argued in \cite{JPG2026} and \cite{renmod}, the variation of $\kappa$ is reflected in the following (exact) relations:\footnote{
We point out that this relation is true for the usual Borel transforms $B[\cD_q](u;\kappa)$ and $B[\cF_q](u;\kappa)$ only in the one-loop approximation. However, for the modified Borel transforms $\tB$ this relation is exact, i.e., to any loop level.}
\bea
\tB[\cD_q](u;\kappa) &=& \exp(u \ln \kappa) \tB[\cD_q](u);
\qquad
\tB[\cF_q](u;\kappa) = \exp(u \ln \kappa) \tB[\cF_q](u).
\label{RScldep} \eea
This suggests that we should expect (even for $\kappa=1$) in the expression for $\tB[\cD_q](u)$ an overall exponential factor $\exp(\tcK u)$. Therefore, we write as our ansatz
\bea
\tB[\cD_q](u) & = &
\exp(\tcK u) \left[ \frac{\tcK_{1/2}^{(0)}}{(1/2-u)^{1}} + \tcK^{(1)}_{1/2} \ln (1 -2 u) + \tcK_{-1} \left( \frac{1}{(1+u)^2} - \frac{1}{(1+u)} \right) \right],
\label{tBDans} \eea
which is an expression with four parameters. These four parameters are then determined by the knowledge of the first four expansion coefficients $({\bar f}_q)_n$ ($n=0,1,2,3$) of $\cF_q$ [$\Leftrightarrow (f_q)_n(\kappa)$], cf.~Eqs.~(\ref{barfj}).\footnote{Later in the paper we will point out that the overall factor $\exp(\tcK u)$ in the ansatz Eq.~(\ref{tBDans}) ensures that the resummation is completely (exactly) independent of the renormalisation scale.}

The characteristic function $F_{\cD_q}(t)$ that enters in the integration that resums the quantities $\cD_q$ and $\cF_q$ is then the inverse Mellin transformation of the modified Borel $\tB[\cD_q](u)$
\be
F_{\cD_q}(t) = \frac{1}{2 \pi i} \int_{u_0 - i \infty}^{u_0 + \infty} du \tB[\cD_q](u) t^u {\big |}_{\tcK \mapsto 0}.
\label{FD1} \ee
This then leads to the following resummation expression for the spacelike quantity $\cD_q(Q^2)$:
\bes
\label{resDqab}
\bea
\cD_q(Q^2) &=& \int_0^{\infty} \frac{dt}{t} F_{\cD_q}(t) \A(t e^{-\tcK} Q^2)
\label{resDqa}
\\
&=& {\Bigg \{} \tcK_{1/2}^{(0)} \int_0^1 \frac{dt}{t} t^{1/2}  \A(t e^{-\tcK} Q^2) + \tcK_{1/2}^{(1)} \int_0^1 \frac{dt}{t} \frac{t^{1/2}}{\ln t}  \left[ \A(t e^{-\tcK} Q^2) - \A(e^{-\tcK} Q^2) \right]
\nonumber\\ &&
+  \tcK_{-1} \int_1^{\infty} \frac{dt}{t} \left( \frac{\ln t}{t} - \frac{1}{t} \right) \A(t e^{-\tcK} Q^2) {\Bigg \}},
\label{resDq}
\eea \ees
where $\A(Q^2)$ is an IR-safe spacelike running QCD coupling that practically coincides with the underlying perturbative coupling $a(Q^2)$ at $|Q^2| > \Lambda_{\rm QCD}^2$.\footnote{Such a coupling $\A(Q^2)$ is also holomorphic, i.e., has no Landau singularities; the only singularities can be on the negative semiaxis $Q^2 <0$. Later we will apply a specific, 3$\delta$AQCD coupling, that has singularities for $Q^2 \leq - M_1^2$ where $M_1^2 \sim m_{\pi}^2$.}
There is also a subleading term in the resummation (\ref{resDq}), $\propto \tcK_{1/2}^{(1)}$, that contains a subtraction $[ \A(t e^{-\tcK} Q^2) - \A(e^{-\tcK} Q^2)]$. 
We refer to \cite{renmod,JPG2026} for details, also App.~B in \cite{JPG2026}.

The resummation for the corresponding timelike quantity $\cF_q(\sigma)$ is then obtained by simply replacing the spacelike coupling $\A$ by the associated timelike coupling $\mA$
\bea
\cF_q(\sigma) &=& {\Bigg \{} \tcK_{1/2}^{(0)} \int_0^1 \frac{dt}{t} t^{1/2}  \mA(t e^{-\tcK} \sigma) + \tcK_{1/2}^{(1)} \int_0^1 \frac{dt}{t} \frac{t^{1/2}}{\ln t}  \left[ \mA(t e^{-\tcK} \sigma) - \mA(e^{-\tcK} \sigma) \right]
\nonumber\\ &&
+  \tcK_{-1} \int_1^{\infty} \frac{dt}{t} \left( \frac{\ln t}{t} - \frac{1}{t} \right) \mA(t e^{-\tcK} \sigma) {\Bigg \}},
\label{resFq}
\eea
where
\be
\mA(\kappa \sigma) = \frac{1}{2 \pi} \int_{-\pi}^{\pi} d \phi \; \A(\kappa \sigma e^{i \phi}).
\label{tH} \ee
If we employed in the evaluation of $\cD_q(Q^2)$, Eq.~(\ref{resDq}), the running perturbative QCD (pQCD) coupling $a(t e^{-\tcK} Q^2)$, the integration would be ambiguous due to the Landau singularities of that coupling (at low positive $t e^{- \tcK} Q^2$), leading to an internal inconsistency of the framework [we note that $\cD_q(Q^2)$ is formally an observable, too]. Therefore, the change $a \mapsto \A$ in Eq.~(\ref{resDq}) is theoretically preferred, and consequently the change $\ma \mapsto \mA$ in Eq.~(\ref{resFq}).\footnote{We denote as $\ma (\kappa \sigma)$ the pQCD timelike coupling, i.e., the coupling obtained when we replace $\A$ by $a$ in the contour integral in Eq.~(\ref{tH}). In \cite{JPG2026}, we used for the timelike couplings instead of $\ma$ and $\mA$, somewhat unfortunately, the notations $\mh$ and $\mH$.}

As in Ref.~\cite{JPG2026}, we will employ for $\A(Q^2)$ the 3$\delta$AQCD coupling \cite{3dAQCD,amu3dAQCD}, i.e., the coupling based on a spectral (discontinuity) function $\rho_{\A}(\sigma) \equiv {\rm Im} \A(Q^2=-\sigma - i \varepsilon)$ that at high $\sigma$ ($> 1 \ {\rm GeV}^2$) coincides with the underlying pQCD discontinuity function $\rho_a(\sigma) \equiv {\rm Im} a(Q^2=-\sigma - i \varepsilon)$ and at low $\sigma$ ($0 < \sigma \lesssim 1 \ {\rm GeV}^2$) it is parametrised by three Dirac delta functions
\be
\rho_{\A}(\sigma) = \pi \sum_{j=1}^3 {\cF}_j \delta(\sigma - M_j^2) + \Theta(\sigma - M_0^2) \rho_{a}(\sigma),
\label{rhoA} \ee
where $\Theta$ is the Heaviside step function, and the squared masses have the following hierarchy: $0 < M_1^2 < M_2^2 < M_3^2 < M_0^2$. Here, $M_1^2 \sim m_{\pi}^2$ is the IR threshold parameter of the coupling $\A$. It is straightforward to show that the corresponding spacelike and timelike couplings are
\bes
\label{AH}
\bea
\A(Q^2) & = & \frac{1}{\pi} \int_{-M^2_{\rm thr}- \eta}^{\infty} \frac{d \sigma \; \rho_{\A}(\sigma)}{(\sigma + Q^2)} \qquad (\eta \to +0),
\label{A} \\
\mA(\sigma) & = & 
\sum_{j=1}^3 \frac{\cF_j}{M_j^2}
\Theta (M_j^2 -\sigma)
+ \frac{1}{\pi} \int_{\Theta(M_0^2-\sigma) \ln(M_0^2/\sigma)}^{\infty} dw \; \rho_a(\sigma e^w),
\label{H} \eea \ees
where $\sigma > 0$.\footnote{
In Ref.~\cite{JPG2026}, we had $\nu_0 \not= 1$, and thus had $\tA_{\nu_0}(Q^2)$ and $\tmA_{\nu_0}(\sigma)$ instead. When $\nu_0 \mapsto 1$, we have the reduction $\tA_{\nu_0} \mapsto \A$ and $\tmA_{\nu_0} \mapsto \mA$. We note that in the expression (81) in \cite{JPG2026}, $\kappa \mapsto 1$ must be set on the left-hand side.}
We refer to Sec.~IV of Ref.~\cite{JPG2026} for the details of the determination and for the resulting values of the parameters ${\cal F}_j$ and $M_j^2$ in these couplings, and we mention that the underlying pQCD coupling $a$, and the corresponding discontinuity function $\rho_a$, are in the (four-loop) lattice-related Lambert MiniMOM (LMM) scheme \cite{MiniMOM}.\footnote{
Perturbative parameters of the MiniMOM (MM) were calculated in \cite{MiniMOM}, including the value of the ${\bar \Lambda}_{\rm MM}$ scaling parameter. The Lambert MiniMOM (LMM) scheme has the scaling parameter ${\bar \Lambda}$ in the coupling the same as in $\MSbar$, while the $\beta$ scheme coefficients $\beta_j$ ($j \geq 2$) are those of MiniMOM, cf.~\cite{3dAQCD}.}
Therefore, we point out that we have to use the expansion coefficients $(f_q)_n$ [that appear in the expansion (\ref{Fqexpgen})] in that LMM scheme. The resulting values of these coefficients (for $n_f=3$ and $\kappa=1$) in the $\MSbar$ and in the LMM schemes are given in Table \ref{tabfj}. The corresponding values of the parameters appearing in the modified Borel transform (\ref{tBDans}), in that scheme, are then deduced and are given in Table \ref{tabtcK}.
\begin{table}
  \caption{The known expansion coefficients $(f_q)_n$ ($0 \leq n \leq 3$) appearing in Eq.~(\ref{Fqexpgen}), for $n_f=3$ and $\kappa(\equiv \mu^2/\sigma)=1$: in the $\MSbar$ scheme, and in the Lambert MiniMOM (LMM) scheme. The corresponding beta-function scheme coefficients $c_j \equiv \beta_j/\beta_0$ are also given.}
\label{tabfj}
\begin{ruledtabular}
\begin{tabular}{r|rrrr|rr}
 scheme & $(f_q)_0$ & $(f_q)_1$ & $(f_q)_2$ & $(f_q)_3$ & $c_2$ & $c_3$  \\
\hline
$\MSbar$ & 1 & 7.73947 & 87.2251 & 1273.9 & 4.47106 & 20.9902 \\
LMM & 1  & 7.73947 & 82.3991 & 1173.97 & 9.29703 & 71.4538 \\
\end{tabular}
\end{ruledtabular}
\end{table}
\begin{table}
  \caption{The parameters $\tcK_q^{(1)}$ appearing in the modified Borel transform $\tB[\cD_q](u)$ Eq.~(\ref{tBDans}), for $n_f=3$ (in the LMM scheme, and for $\kappa=1$).}
\label{tabtcK}
\begin{ruledtabular}
\begin{tabular}{r|rrr}
 $\tcK$ & $\tcK_{1/2}^{(0)}$ & $\tcK_{1/2}^{(1)}$ & $\tcK_{-1}$ \\
\hline
1.59242  & 0.5  & 0.015126 & 0.122401 \\
\end{tabular}
\end{ruledtabular}
\end{table}

The coupling $\A(Q^2)$ practically coincides with the underlying pQCD coupling $a(Q^2)$ at $|Q^2| > 1 \ {\rm GeV}^2$,
namely $\A(Q^2) - a(Q^2) \sim (\Lambda_{\rm QCD}^2/Q^2)^5$. In the deep IR range we have $\A(Q^2) \sim Q^2$ ($|Q^2| < 0.1 \ {\rm GeV}^2$), as suggested by large-volume lattice QCD calculations \cite{3dLVLattice,Lattb,Latt2b,Latt2c}.

The entire resummation procedure is completely (exactly) independent of the renormalisation scale $\mu^2$, as already shown in our previous works \cite{renmod,PLB848,NPB1007,JPG2026}. We can check (for example, numerically) the following: If we start our procedure \text{black}{based on the first four known coefficients of the perturbation series} Eq.~(\ref{Fqexpgen}) with a specific $\kappa \not= 1$ [i.e., on the basis of $(f_q)_n(\kappa)$, $n=0,1,2,3$], and using the same ansatz (\ref{tBDans}) where now the four parameters $\tcK_j$ are $\kappa$-dependent, we obtain the very same values of these parameters $\tcK_{1/2}^{(0)}$, $\tcK_{1/2}^{(1)}$ and $\tcK_{-1}$ as in the case $\kappa=1$, and we obtain (exactly) $\tcK(\kappa)=\tcK + \ln \kappa$. This is consistent with the (exact) relations Eq.~(\ref{RScldep}). The characteristic function $F_{\cD_q}(t)$ is then the same as in the case $\kappa=1$, cf.~Eq.~(\ref{FD1}). Therefore, in the resummation (\ref{resFq}) nothing changes, except possibly the argument $\exp(-\tcK) \sigma$ in the (timelike) running couplings $\mA(t e^{-\tcK} \sigma)$ and $\mA(e^{-\tcK} \sigma)$ in the integrals there. However, in the approach with $\kappa \not= 1$, the factor $\sigma$ changes everywhere to $\kappa \sigma$ [cf.~the expansion (\ref{Fqexpgen})], and the factor $\exp(-\tcK(\kappa))$ changes to
\be
\exp(-\tcK(\kappa)) = \exp(-\tcK - \ln \kappa) = \exp(-\tcK) \times \frac{1}{\kappa}.
\label{exptK} \ee
This implies that the argument $\exp(-\tcK) \sigma$ remains (exactly) invariant under the change of the renormalisation scale (i.e., under the change $\kappa =1$ to $\kappa \not= 1$). Therefore, our resummation procedure Eq.~(\ref{resFq}) [and Eq.~(\ref{resDq})] is completely independent of the variation of the renormalisation scale $\mu^2 = \kappa \sigma$.

\begin{figure}[htb] %\unitlength=1mm
\begin{minipage}[b]{.49\linewidth}
\includegraphics[width=80mm,height=50mm]{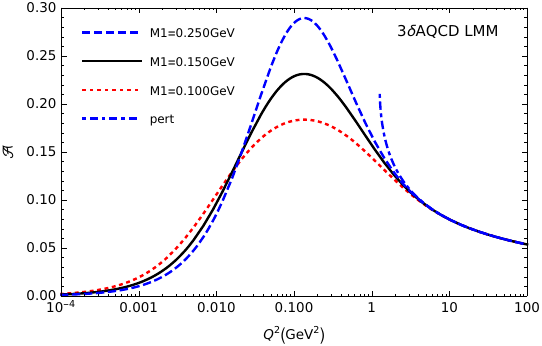}
\end{minipage}
\begin{minipage}[b]{.49\linewidth}
  \includegraphics[width=80mm,height=50mm]{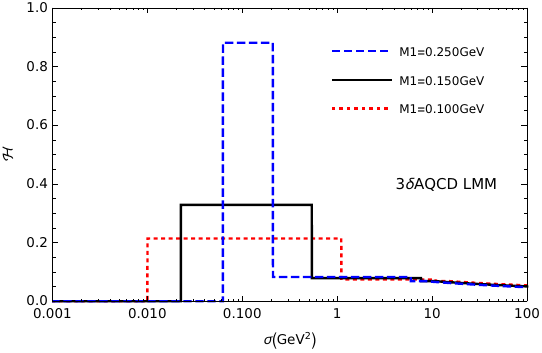}
\end{minipage} \vspace{4pt}
\caption{\footnotesize The spacelike running coupling $\A(Q^2)$ for positive $Q^2$ (left-hand figure), and the timelike running coupling $\mA(\sigma)$ (right-hand figure), in 3$\delta$AQCD, with $n_f=3$, $\alpha_s^{\MSbar}(M_Z^2)=0.1180$, and three different values of the IR-threshold scale parameter $M_1$.} 
\label{FigAH}
\end{figure}

In Figs.~\ref{FigAH} we depict the spacelike and timelike couplings, $\A(Q^2)$ and $\mA(\sigma)$, for three different values of the threshold parameter $M_1$. The underlying pQCD coupling is at $n_f=3$ and corresponds to the world average value $\alpha_s^{\MSbar}(M_Z^2)=0.1180$.
We mention that the authors of \cite{Pelaez}, within the framework of the Curci-Ferrari version of QCD, obtained the ghost-gluon coupling that has a similar form as 3$\delta$AQCD $\A(Q^2)$ here.

For the evaluation of the quark pole mass parameter quantity $\cF_q(\sigma)$, Eq.~(\ref{resFq}), we use for the $\MSbar$ masses $\bam_q$ [$\equiv \bam_q(\bam_q^2)$] the PDG values \cite{PDG2025}: $\bam_b=(4.183 \pm 0.007)$ GeV and $\bam_c=(1.2730 \pm 0.0046)$ GeV. This then leads to the following results for the pole mass values $m_q \equiv m_q(\sigma=\bam_q^2)$ in our approach: 
\begingroup\color{black}
\bes
\label{mqp}
\bea
m_b & = & \left[ 5.008 \pm 0.089(\A) ^{+0.057}_{-0.051}(M_1) ^{+0.028}_{-0.027}(\alpha_s) ^{+0.008}_{-0.007} (\bam_b)  \pm 0.002 ({\rm an}) \right] \ {\rm GeV} = 5.008(110) \ {\rm GeV},
\label{mbp} \\
m_c & = &   \left[ 1.670 \pm 0.098(\A)^{+0.156}_{-0.108}(M_1) ^{+0.014}_{-0.015}(\alpha_s) ^{+0.006}_{-0.007}(\bam_c) \pm 0.039({\rm an})\right] \ {\rm GeV} = 1.670(189) \ {\rm GeV}.
\label{mcp} \eea \ees
\endgroup
At the end, various uncertainties were then added in quadrature.

The uncertainties '$(M_1)$' originate from varying the IR-threshold parameter of the coupling: $M_1 = (0.150^{+0.100}_{-0.050})$ GeV; and the uncertainties '$(\alpha_s)$' from varying the world average value \cite{PDG2025} of $\alpha_s$: $\alpha_s^{\MSbar}(M_Z^2)=0.1180 \pm 0.0009$.

\begingroup\color{black}
The uncertainties '(an)' come from varying the ansatz for $\tB[\cD_q](u)$. We recall that for the central value, the ansatz was taken according to Eq.~(\ref{tBDans}). For '(an)' uncertainties, we used two other ans\"atze: 1.) the NLO $u=1/2$ IR term $\sim \ln(1 - 2 u)$ was not included, but the (LO) $u=3/2$ IR term $\sim 1/(3/2-u)$ was included; 2.) the UV $u=-1$ term(s) were not included, but the $u=3/2$ IR term $\sim 1/(3/2-u)$ was included. This latter choice increased $m_c$ by $0.039$ GeV, while the first choice increased $m_b$ by 0.002 GeV. The value of $m_b$ remains remarkably stable under these variations of the ansatz.  

The uncertainties '$(\A)$', which are the dominant ones for $m_b$, are uncertainties related with the variation of the AQCD coupling framework, and to a lesser degree, to the variation of the renormalisation scheme dependence. They were obtained in the following way. We notice that the maximum values of the 3$\delta$AQCD coupling $\A(Q^2)$ (for $Q^2>0$) are $\A_{\rm max} = 0.2312^{+0.0582}_{-0.0476}$, for the IR threshold values $M_1=0.150^{+0.100}_{-0.050}$ GeV, cf.~the left-hand Fig.~\ref{FigAH}. We then used 2$\delta$AQCD coupling instead\footnote{i.e., whose spectral function $\rho_{\A}(\sigma)$, cf.~Eq.~(\ref{rhoA}), has two (instead of three) Dirac delta terms.}, either in the same LMM scheme, or in the $\MSbar$ scheme. The resulting couplings $\A(Q^2)$ have their maximum value (for $Q^2 \geq 0$) at $Q^2=0$. We chose those values to coincide with the mentioned $\A_{\rm max}$ values of the 3$\delta$AQCD model, i.e., $\A(0) (=\A_{\rm max})=0.2312^{+0.0582}_{-0.0476}$. In this way we extracted different values of $m_b$ and $m_c$. The largest deviation from our central value $m_b=5.008$ GeV of Eq.~(\ref{mbp}) was for 2$\delta$AQCD in the LMM scheme with $\A(0)=0.2312+0.0582$ ($=0.2894$), namely $+0.089$ GeV. And the largest deviation  from our central value $m_c=1.670$ GeV of Eq.~(\ref{mcp}) was for 2$\delta$AQCD in the $\MSbar$ scheme with $\A(0)=0.2312-0.0476$ ($=0.1836$), namely $-0.098$ GeV. Hence, we regard as '$(\A)$' uncertainties $\pm 0.089$ for $m_b$, and $\pm 0.098$ for $m_c$.\footnote{\textcolor{black}{The case of 2$\delta$AQCD in the LMM scheme and with $\A(0)=0.2894$ gives a similar, but positive, deviation of $m_c$, namely $+0.079$.}}

We point out that 2$\delta$AQCD couplings, by construction, have the same de facto coincidence in the UV regime as 3$\delta$AQCD, namely $\A(Q^2) - a(Q^2) \sim (\Lambda_{\rm QCD}^2/Q^2)^5$, and one (IR) parameter remains free, namely the value of $\A(0)$, and that latter value was fixed by the results of 3$\delta$AQCD coupling. In this context, we see that the 3$\delta$AQCD model is more natural, in the sense that the ranges of the (IR) threshold scale parameters of the spectral function $\rho_{\A}(\sigma)$ of the coupling $M_1$ are in the (naively) expected range of the lowest hadronic mass scales of QCD: $M_1 \sim m_{\pi}$ ($M_1=0.150^{+0.100}_{-0.050}$ GeV). The 2$\delta$AQCD couplings that have comparably large values $\A_{\rm max}$ ($= 0.2312^{+0.0582}_{-0.0476}$) have the parameters $M_1$ significantly larger (in the range $0.65$-$1.23$ GeV). It is not surprising that the 3$\delta$AQCD is better in the aforementioned sense, because the 3$\delta$AQCD coupling follows in the IR sector closely the coupling determined as a product of the gluon and ghost Landau-gauge dressing functions whose values are largely known in the IR regime by the large-volume lattice calculations, see e.g. \cite{3dLVLattice}.\footnote{\textcolor{black}{In 3$\delta$QCD, when $Q^2 \to 0$, we have $\A(Q^2) \sim Q^2 \to 0$. The maximum of $\A(Q^2)$ for $Q^2 \geq 0$ is achieved (in the LMM scheme) at $Q^2_0=0.135 \ {\rm GeV}^2$.}}
It is these lattice values that largely determine the behaviour of $\A(Q^2)$ in the IR range, while there are no clear clues as to how the 2$\delta$AQCD coupling behaves in this range.

In Figs.~\ref{FigAH2d} we present the spacelike and timelike couplings of the 2$\delta$AQCD framework in the LMM scheme, with the mentioned maximal values $\A(0)=0.2312^{+0.0582}_{-0.0476}$. These Figures can be compared to Figs.~\ref{FigAH} of the 3$\delta$AQCD framework. Further, the curves for 2$\delta$AQCD in the $\MSbar$ scheme are similar to those in the LMM scheme presented in Figs.~\ref{FigAH2d}.
\begin{figure}[htb] %\unitlength=1mm
\begin{minipage}[b]{.49\linewidth}
\includegraphics[width=80mm,height=50mm]{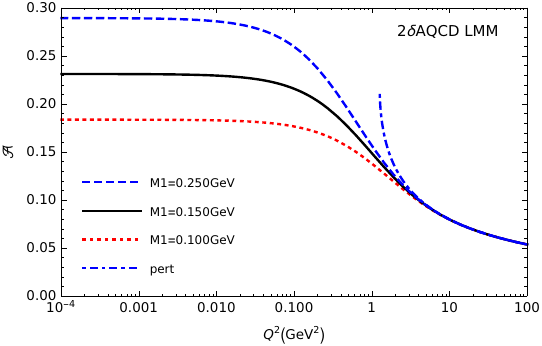}
\end{minipage}
\begin{minipage}[b]{.49\linewidth}
  \includegraphics[width=80mm,height=50mm]{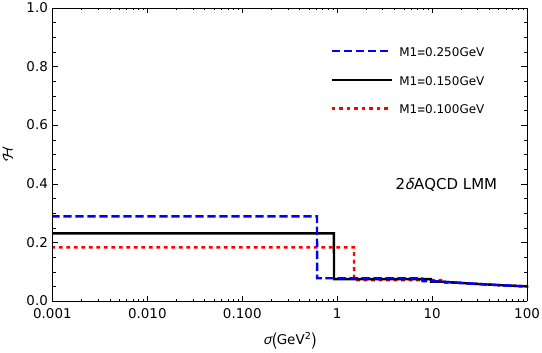}
\end{minipage} \vspace{4pt}
\caption{\footnotesize \textcolor{black}{As Figs.~\ref{FigAH}, but now for 2$\delta$AQCD framework in the LMM scheme (see the text for details). Three different couplings are presented, with the maximal values $\A_{\rm max}=0.2312^{+0.0582}_{-0.0476}$ ($=\A(0)$) that correspond to the maximal values of 3$\delta$AQCD with the IR thresholds $M_1=0.150^{+0.100}_{-0.050}$ GeV (in 3$\delta$AQCD), cf.~the left-hand Figure \ref{FigAH}. We point out that the IR threshold values ($M_1$) of these 2$\delta$AQCD couplings are significantly higher than $0.150^{+0.100}_{-0.050}$ GeV (namely: $M_1=0.961^{-0.178}_{+0.264}$ GeV).}}
\label{FigAH2d}
\end{figure}
\endgroup

We can also evaluate all the above quantities directly from the corresponding perturbation series (in the $\MSbar$ scheme), by (naively) truncating the series at the smallest term. \textcolor{black}{This naive approach can be regarded as a first approximation to the resummation of a divergent perturbation series that is dominated by IR renormalons, and the smallest term can be regarded as a first estimate of the leading IR renormalon ambiguity \cite{Renormalons}.} 
It turns out that in the perturbation series (\ref{Fqexp}) for $m_b$ (i.e., when $\sigma=\bam_b^2$, $n_f=3$, $\kappa=1$) the terms $({\bar f}_q)_n {\bar a}(\bam_b^2)^{n+1}$ are the smallest and almost equal for $n=2$ and $n=3$; therefore, we truncate with half of the term $n=3$, i.e., $(1/2) ({\bar f}_q)_3({\bar a}\bam_b^2)^{4}$ \textcolor{black}{as the last term}, and the truncation uncertainty is taken as $\pm  ({\bar f}_q)_3{\bar a}(\bam_b^2)^{4}$. On the other hand, in the series (\ref{Fqexp}) for $m_c$ (i.e., when $\sigma=\bam_c^2$, $n_f=3$, $\kappa=1$) the terms $({\bar f}_q)_n {\bar a}(\bam_c^2)^{n+1}$ are the smallest and almost equal for $n=0$ and $n=1$; therefore, we truncate with half of the term $n=1$, i.e., $(1/2) ({\bar f}_q)_1{\bar a}(\bam_c^2)^{2}$, and the truncation uncertainty is taken as $\pm  ({\bar f}_q)_1{\bar a}(\bam_c^2)^{2}$. This then leads to the following results for the pole masses in the \textcolor{black}{truncation perturbation series (TPS)} approach:
\bes
\label{mqpT}
\bea
(m_b)_{\rm (TPS)} & = & \left[ 5.004 \pm 0.157 ({\rm TPS}_{mb})  ^{+0.023}_{-0.022}(\alpha_s) \pm 0.007 (\bam_b)  \right] \ {\rm GeV},
\label{mbpT} \\
(m_c)_{\rm (TPS)} & = &   \left[ 1.585 \pm 0.202 ({\rm TPS}_{mc}) ^{+0.013}_{-0.012}(\alpha_s) \pm 0.005(\bam_c) \right] \ {\rm GeV}.
\label{mcpT} \eea \ees
Comparison with the results Eqs.~(\ref{mqp}), of the renormalon-resummed approach, shows that the truncation uncertainties in Eqs.~(\ref{mqpT}) are in general significantly higher than the IR-threshold uncertainties '($M_1$)' in Eqs.~(\ref{mqp}). We point out that the truncation uncertainties '(${\rm TPS}_{mq}$)' mainly reflect the principal IR-renormalon ambiguities which, in the resummed approach using the 3$\delta$AQCD coupling, would correspond approximately to the uncertainties '($M_1$)' originating from the variation of the IR-threshold parameter $M_1$ of the coupling 3$\delta$AQCD coupling in an expected range ($M_1 \sim m_{\pi}$). In Eqs.~(\ref{mqpT}) we did not include variations originating from the variation of the renormalisation scale $\mu^2= \kappa \bam^2_q$, but assumed that they are, at least partly, contained in the truncation uncertainties '(${\rm TPS}_{mq}$)'. \textcolor{black}{Namely, if the perturbation series is dominated by an IR renormalon, it is straightforward to show that the variation of the renormalisation scale of a TPS of that series (by a factor $\kappa \sim 1$) gives us a variation of that TPS that is of the order of the next term beyond that TPS. The latter term is, in our case, roughly equal to the minimal term in the series, which, in turn, is of the order of the IR renormalon ambiguity \cite{Renormalons}.}

\begin{figure}[htb] %\unitlength=1mm
\centering\includegraphics[width=80mm]{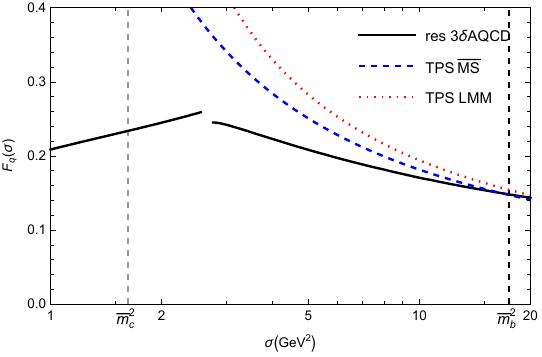}
 \vspace{0pt}
 \caption{\footnotesize The function $F_q(\sigma)$, for $n_f=3$, as a function of $\sigma$: the resummed version Eq.~(\ref{resFq}). For comparison, the truncated series (TPS) in the $\MSbar$ and the LMM schemes are included. See the text for details. The relevant values are for $\sigma=\bam_c^2$ ($=1.6205 \pm  0.0117 \ {\rm GeV}^2$) and for $\sigma=\bam_b^2$ ($=17.4975 \pm 0.0586 \ {\rm GeV}^2$), leading to the values of pole mass $m_b$ and $m_c$ via the relations (\ref{Fqexp})-(\ref{mqpole}). The small (steplike) discontinuity of the curve for the resummed values is at $\sigma= e^{\tcK} M_2^2$ ($ \approx 2.66 \ {\rm GeV}^2$), an artifact of our parametrisation of the spectral function $\rho_{\A}(\sigma)$ in the low-$\sigma$ range, Eqs.~(\ref{rhoA}) (see the text for more explanation).}
 \label{FigFqsig}
\end{figure}

In Fig.~\ref{FigFqsig} we present the quantity $\cF_q(\sigma)$, Eq.~(\ref{Fqexp}), as a function of $\sigma$ in the $\sigma$-interval of interest: the resummed quantity [cf.~Eq.~(\ref{resFq})], for the central input values of the 3$\delta$QCD $n_f=3$ couplings $\A$ and $\mA$: $M_1=0.150$ GeV and $\alpha_s^{\MSbar}(M_Z^2)=0.1180$. For comparison, we include in Fig.~\ref{FigFqsig} the quantity $\cF_q(\sigma)$ evaluated as a TPS in the $\MSbar$ and LMM schemes, where we truncate at the term $\sim a^3$ and add half of the term $\sim a^4$.\footnote{This is the truncation of $\cF_q(\sigma)$ we used and discussed when $\sigma=\bam_b^2$, cf.~Eq.~(\ref{mbpT}).}
The curve for the resummed values has a small steplike discontinuity at $\sigma= e^{\tcK} M_2^2$ ($ \approx 2.66 \ {\rm GeV}^2$), which is an artifact of the used parametrisation of the spectral function $\rho_{\A}(\sigma)$ in the low-$\sigma$ range via the Dirac delta functions, cf.~Eqs.~(\ref{rhoA}) and (\ref{H}). The curve has also similar small steplike discontinuities at $\sigma= e^{\tcK} M_j^2$ ($j=1,3,0$), but these are outside the depicted $\sigma$-range ($1. \ {\rm GeV}^2 \leq \sigma \leq 20. \ {\rm GeV}^2$) relevant for our quantities $m_c$ and $m_b$. We note that the mentioned steplike discontinuities of $F_q(\sigma)$ come from the (small) IR-subleading part of the resummation proportional to $\tcK^{(1)}_{1/2}$ in Eq.~(\ref{resFq}). 

\begin{figure}[htb] %\unitlength=1mm
\begin{minipage}[b]{.49\linewidth}
\includegraphics[width=80mm,height=50mm]{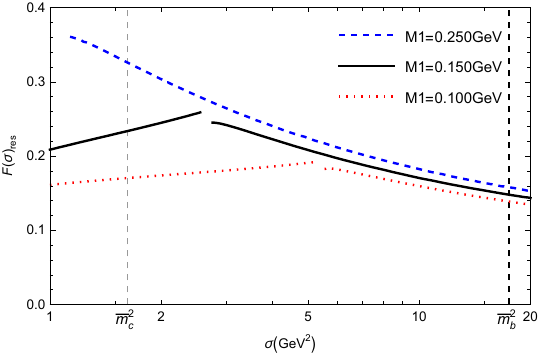}
\end{minipage}
\begin{minipage}[b]{.49\linewidth}
  \includegraphics[width=80mm,height=50mm]{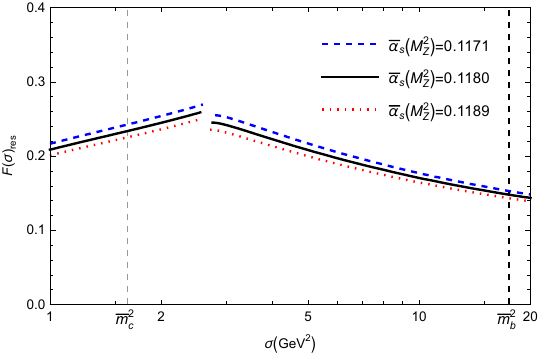}
\end{minipage} \vspace{0pt}
\caption{\footnotesize The resummed values of $\cF_q(\sigma)$ as in Fig.~\ref{FigFqsig}: (a) when the IR-threshold scale $M_1$ is varied and $\alpha_s^{\MSbar}(M_Z^2)=0.1180$; (b) when $\alpha_s^{\MSbar}(M_z^2)$ value is varied and $M_1=0.150$ GeV.}
\label{FigFqsigM1Al}
\end{figure}

In Figs.~\ref{FigFqsigM1Al} we present again the resummed values of $F_q(\sigma)$, for various IR-threshold parameter values $M_1=(0.150^{+0.100}_{-0.050})$ GeV (left-hand side) and for various values of $\alpha_s^{\MSbar}(M_z^2)=0.1180 \pm 0.009$.
Again, the small steplike discontinuities appear at the corresponding values $\sigma=e^{\tcK} M_2^2$.

\section{HQET and determination of nonperturbative parameters from heavy-meson masses}
\label{sec:HQETdet}
  
\subsection{HQET operators and matrix elements}
\label{sec:HQET}

Within the Heavy Quark Effective Theory (HQET), one can effectively describe the dynamics of a heavy meson composed of the static heavy-quark field $h_v$ with four-velocity $v$ and a light quark $\ell$.
The operators of $O(1/m_h)$\footnote{Here, $m_h$ is the pole mass of the heavy quark.} in the OPE are the kinetic-energy operator and chromomagnetic operator, which are defined as 
\bes
\label{HQET}
\bea
O_{\rm kin} &=& - \bar h_v (i D_\perp)^2 h_v,
  \qquad D_\perp^\mu \equiv D^\mu - v^\mu (v\!\cdot\! D),
\label{HQETa}\\   
O_{\rm cm}  &=& \frac{g_s}{2}\,\bar h_v\,\sigma_{\mu\nu}G^{\mu\nu}\,h_v,
  \qquad \sigma_{\mu\nu}=\frac{i}{2}[\gamma_\mu,\gamma_\nu]. 
\label{HQETb}   
\eea \ees

Reparameterization invariance implies that the Wilson coefficient of $O_{\rm kin}$
is exactly $1$ for all orders\cite{Luke:1992cs}, while the chromomagnetic coefficient $C(\mu)$
is obtained by matching QCD to HQET. This $C(\mu)$ was calculated to two loops in \cite{ABN, Czarnecki:1997dz} and to three loops in \cite{Grozin:2007fh} together with their corresponding anomalous dimension.

For a heavy meson at rest, the standard nonperturbative parameters of HQET are defined through the
matrix elements (in the infinite-mass limit):
\begin{eqnarray}
  \mu_\pi^2 &\equiv&
  -\langle H|\,\bar h (iD_\perp)^2 h\,|H\rangle, \qquad
  \mu_G^2 \equiv
  -\left\langle H\left|\,\bar h\frac{g_s}{2}\sigma_{\mu\nu}G^{\mu\nu} h
  \right|H\right\rangle.
\label{mu2s}
\end{eqnarray}
These are heavy-quark symmetry universal (same for $H=B,D$ up to $1/m_h$ effects).

The mass of a heavy--light ground-state meson $H$ within HQET with spin $s=0$ (pseudoscalar, $H$)
or $s=1$ (vector, $H^*$), yields 
\begin{equation}
  M_H^{(s)} =
  m_h
  + \bar\Lambda
  + \frac{\mu_\pi^2}{2\,m_h}
  + A(s)\,C(m_h^2)\,\frac{\mu_G^2(m_h^2)}{2\,m_h}
  + \mathcal{O}\!\left(\frac{\Lambda_{\rm QCD}^3}{m_h^2}\right),
  \label{eq:MHexp}
\end{equation}
where the spin weight is $A(0)=-1,\ A(1)=\frac{1}{3}$.

Thanks to the structure of the heavy mass given in (\ref{eq:MHexp}), we can use the spin structure to isolate the NP parameter that we want to determine. For instance, as $A(0)+3A(1)=0$, we can define define the spin-averaged mass $\overline M_H \equiv (M_H^{(0)}+3M_H^{(1)})/4$ and therefore the chromomagnetic term cancels at order $1/m_h$ and one gets
\begin{equation}
  \overline M_H =
  m_h
  + \bar\Lambda
  + \frac{\mu_\pi^2}{2\,m_h}
  + \mathcal{O}\!\left(\frac{\Lambda_{\rm QCD}^3}{m_h^2}\right).
  \label{eq:spinave}
\end{equation}

Another alternative is to consider the hyperfine splitting, since in this case, we can isolate the $\mu_G^2$ parameter with reduced explicit $1/m_h$ dependence, giving
\begin{equation}
 M_{H^*}^2 - M_H^2  = \frac{4}{3} C(m_h^2)\,\mu_G^2(m_h^2)\left( 1- \Lambda_{\rm eff}\frac{1}{m_h}\right)
  + \mathcal{O}\!\left(\frac{\Lambda_{\rm QCD}^4}{m_h^2}\right),
  \label{eq:hypersq}
\end{equation}
where $\Lambda_{\rm eff}$ is a combination of NP parameters that parametrize the spin-orbit interaction, and bilocal elements of chromomagnetic interactions \cite{GandN}.

Another important NP parameter has been introduced in eq.(\ref{eq:MHexp}), which is $\bar\Lambda$. This contribution appears at $\mathcal{O}(\Lambda_{\rm QCD})$ of light degrees of freedom in the static limit. A common definition is \cite{Hayashi:2021vdq}
\begin{equation}
  \bar\Lambda \equiv \lim_{m_h\to\infty}\left(M_H - m_h\right),
\label{barL}
\end{equation}
Based on this definition, $\bar\Lambda$ is ambiguous, as one must specify how the heavy quark pole mass $m_h$ is defined and/or regularised.

The pole mass should be IR-finite order by order, but their perturbative sum is ill-defined due
to the renormalons in the Borel plane. To deal with the renormalon ambiguity, we can consider the dominant one (the singularity closest to the origin in the Borel plane) which is given by the Borel parameter $u=1/2$. The perturbative series relating
$m_h$ to any short-distance mass inherits this ambiguity of order
$\mathcal{O}(\Lambda_{\rm QCD})$.
The crucial physical point is that in Eq.~\eqref{eq:MHexp} the $\mathcal{O}(\Lambda_{\rm QCD})$
ambiguity in $m_h$ cancels against the corresponding ambiguity in
$\bar\Lambda$ (and similarly higher-power renormalons cancel against higher-dimension
matrix elements), so that the meson mass is unambiguous \cite{Hayashi:2023axn}.

\subsection{Extracting ${\hat C}(m_q^2)$, $\Lambda_{\rm eff}$, $\bar\Lambda,\mu_\pi^2,{\hat{\mu}}_G^2$ from $B/D$ spectroscopy}
\label{sec:det}

Eqs.~\eqref{eq:spinave} and \eqref{eq:hypersq} suggest a clear strategy to determine the NP parameters $\bar\Lambda,\mu_\pi^2,\mu_G^2$. But before that, we should clarify how $\Lambda_{\rm eff}$ was previously obtained.
Hereafter, we will deal with the renormalization group (RG) invariant  quantities, that correspond to simple renaming 
\bes
\label{Cmu}
\begin{eqnarray}
  C(m_h^2; \mu^2)&=& \hat{C}(m_h^2) K(\mu^2), \qquad
  {\mu}_G^2(\mu^2) = \hat{\mu}_G^2/K(\mu^2),
\label{CmuRScl}\\
\hat{C}(m_h^2) &\equiv& \pi^{\nu_0} \cF_C(m_h^2).
\label{CF}
\end{eqnarray} \ees
This implies that the product $C(m_h^2) \mu_G^2(m_h^2)$ can be rewritten in the RG-invariant form ${\hat C}(m_h^2) \hat{\mu}_G^2$. 
In \cite{JPG2026} we have evaluated the truncated perturbation series (TPS) in powers of the $\MSbar$ pQCD coupling:
\be
\cF_C(\sigma)^{{\rm TPS[N]};\MSbar} = {\bar a}(\sigma)^{\nu_0} + \sum_{j=1}^{N-1} ({\bar f}_C)_j {\bar a}(\sigma)^{\nu_0+j},
\label{FCTPSMSb} \ee
where $\nu_0=1/3$ (for $n_f=3$), and the bars indicate that these are the quantities in the $\MSbar$ scheme.
The perturbation expansion of this quantity in a general scheme and at a general renormalisation scale $\mu^2=\kappa \sigma$ has the form
\be
\cF_C(\sigma) = a(\kappa \sigma)^{\nu_0} + \sum_{j=1}^{N-1} (f_C)_j(\kappa)  a(\kappa \sigma)^{\nu_0+j}.
\label{FCexp} \ee
The resummed version has the form
\be
\cF_C(\sigma)_{\rm res.} = \int_0^{\infty}  \frac{dt}{t}  F^{(C)}_{\cDo}(t) \tmA_{\nu_0}(t e^{-\tK_e^{(1)}} \sigma),
\label{resFC}   \ee
where $F^{(C)}_{\cDo}(t)$ is the characteristic function of an associated spacelike quantity $\cD^{(1)}(Q^2)$, and $\tmA_{\nu_0}(\sigma)$ the timelike analog of the generalised logarithmic derivative coupling $\tA_{\nu_0}(Q^2)$ in 3$\delta$AQCD 
\be
\tmA_{\nu_0}(\sigma) =\frac{\sin (\pi \nu_0) \beta_0^{1-\nu_0}}{\pi^2 (1-\nu_0)}
\left\{ \pi \sum_{j=1}^3 \frac{\cF_j}{M_j^2}
\Theta (M_j^2 -\sigma)
\ln^{1-\nu_0} \left( \frac{M_j^2}{\sigma} \right)
+ \int_{\Theta(M_0^2-\sigma) \ln(M_0^2/\sigma)}^{\infty} dw \; w^{1-\nu_0}  \rho_a(\sigma e^w) \right\}.
%{\rm max}(0,\ln(M_0^2/\sigma))
\label{tHnu0} \ee
We refer to \cite{JPG2026} for details.

In contrast to the previous determination \cite{JPG2026}, we will now take $n_f=3$ even for the $H=B, (m_h=m_b)$ meson case [i.e., not just for the $H=D$ ($m_h=m_c$) case], since the (known) charm mass contribution becomes negligible when $n_f=3$ \cite{AyaPin2026} (\textcolor{black}{the same logic applied to the evaluation of $m_b$}, cf.~also footnote \ref{BBB}).
\begin{figure}[htb] %\unitlength=1mm
\centering\includegraphics[width=80mm]{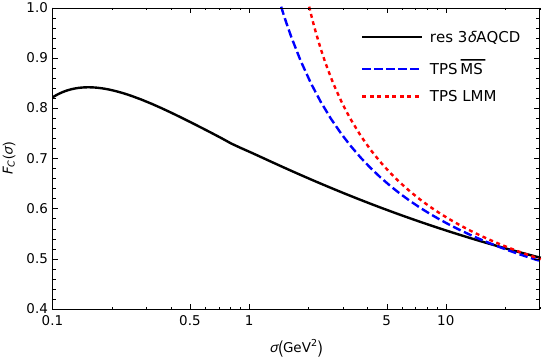}
 \vspace{0pt}
 \caption{\footnotesize The renormalon-resummed $\cF_C(\sigma)$ [$\equiv \pi^{-\nu_0} {\hat C}(\sigma)$] as a function of the squared timelike momentum (squared mass) $\sigma$, in 3$\delta$AQCD, for $n_f=3$, $M_1=0.150$ GeV and $\alpha_s^{\MSbar}(M_Z^2)=0.1180$, and for a wide range $0.1 \ {\rm GeV}^2 \leq \sigma \leq 30 \ {\rm GeV}^2$. For comparison, we include the corresponding pQCD TPS Eq.~(\ref{FCTPSMSb})-(\ref{FCexp}) in the $\MSbar$ and LMM schemes, with four terms included ($N=4$, $\kappa=1$).}
 \label{FigFCsig}
\end{figure}
\begin{figure}[htb] %\unitlength=1mm
\begin{minipage}[b]{.49\linewidth}
\includegraphics[width=80mm,height=50mm]{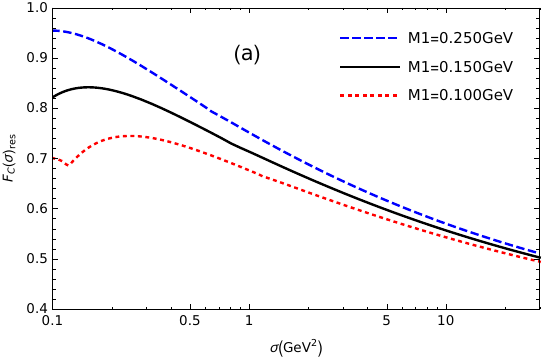}
\end{minipage}
\begin{minipage}[b]{.49\linewidth}
  \includegraphics[width=80mm,height=50mm]{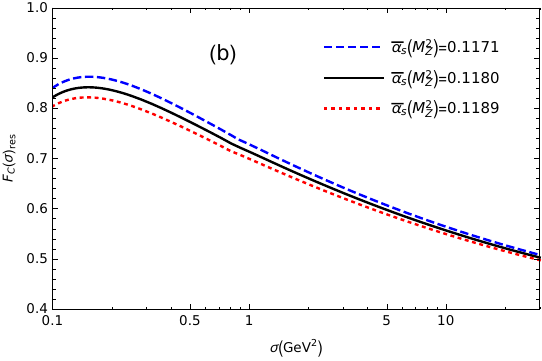}
\end{minipage} \vspace{0pt}
\caption{\footnotesize The resummed values of $\cF_C(\sigma)$ as in Fig.~\ref{FigFCsig}, where now: (a) the IR-threshold scale $M_1$ is varied and $\alpha_s^{\MSbar}(M_Z^2)=0.1180$; (b) the $\alpha_s^{\MSbar}(M_z^2)$ value is varied and $M_1=0.150$ GeV.}
\label{FigFCsigM1Al}
\end{figure}
We present in Figs.~\ref{FigFCsig} and \ref{FigFCsigM1Al} the resummed quantity $\cF_C(\sigma)$ as a function of $\sigma$ in a wide range $0.1 \ {\rm GeV}^2 \leq \sigma \leq 30. \ {\rm GeV}^2$ relevant for our evaluations, for $n_f=3$ and for various values of the input parameters of the 3$\delta$AQCD coupling used: $M_1=(0.150^{+0.100}_{-0.050})$ GeV and $\alpha_s^{\MSbar}(M_Z^2)=0.1180 \pm 0.009$. The relevant values of $\sigma$ are $\sigma=m_c^2, m_b^2$, where these two values were obtained in Eqs.~(\ref{mqp}) with resummation using the same 3$\delta$AQCD couplings.

Now the values obtained in this way are the following:
\begingroup\color{black}
\bes
\label{FCs}
\bea
\cF_C(m_b^2)_{\rm res.} & = & 0.5105 \pm 0.0137(\A)  ^{+0.0077}_{-0.0081}(M_1) \pm 0.0050(\alpha_s) \mp 0.0002(\bam_b) = 0.510(17),
\label{FCmb} \\
\cF_C(m_c^2)_{\rm res.} & = & 0.6364 \pm 0.0356 (\A) ^{+0.0098}_{-0.0163}(M_1) ^{+0.0098}_{-0.0094}(\alpha_s) \mp 0.0006(\bam_c) = 0.636(40).
\label{FCmc}
\eea \ees
We recall that '$(\A)$' uncertainties, as in Eqs.~(\ref{mqp}) for pole masses, originate from the variation of the AQCD framework described in detail there. Specifically, the largest deviation from the central values $\cF_C(m_b^2)_{\rm res.}= 0.5105$ and $\cF_C(m_c^2)_{\rm res.} =0.6364$ are obtained in 2$\delta$AQCD in the LMM scheme with $\A(0)=0.2894$: the extracted values are: $m_b=5.097$ GeV, $m_c=1.749$ GeV, $\cF_C(m_b^2)_{\rm res.}= 0.5242$ (deviates by $+0.0137$ from 0.5105), $\cF_C(m_c^2)_{\rm res.}=0.6720$ (deviates by $+0.0356$ from 0.6364).

While the details of the $u=1/2$ IR renormalon structure of the transforms $\tB[\cF_C](u)$ are known \cite{GandN}, the detailed structure beyond the $u=1/2$ renormalons is not known. Therefore, we have not estimated the '(an)' uncertainties (i.e., those related with the variation of the renormalon ans\"atze) for the quantities $\cF_C(m_q^2)$ and the related quantities.

We also note that for each uncertainty, we are not just changing the corresponding parameter (e.g., $M_1$), but also the values of the pole masses $m_b$ and $m_c$ accordingly, cf.~Eqs.~(\ref{mqp}).

The ratio of these two functions at $m_h=m_c$ and $m_h=m_b$ is then
\be
 \frac{{\hat C}(m_b^2)}{{\hat C}(m_c^2)} = \frac{ \cF_C(m_b^2) }{\cF_C(m_c^2)}
 = 0.8022 \mp 0.0221(\A) ^{-0.0002}_{+0.0080}(M_1) ^{-0.0042}_{+0.0041}(\alpha_s) \mp 0.0002 (\bam_b) \pm 0.0007(\bam_c) = 0.802(24).
  \label{rathatC} \ee
  Here, the uncertainty '$(\A)$' was obtained by evaluating the ratio of the aforementioned 2$\delta$AQCD values, $0.5242/0.6720=0.7801$, and this deviates by $-0.0221$ from $0.8022$.
 \endgroup
 In these ratios, we further took into account that the uncertainties coming from the IR regime ['$(M_1)$'] in the numerator and the denominator are  completely correlated, as are the '$(\alpha_s)$' uncertainties. The various uncertainties were then added in quadrature. It turns out that this ratio is the leading order approximation for the following ratio of mass splitting between the ground-state pseudoscalar and vector mesons, in the bottom and charm quark systems \cite{GandN} (cf.~also \cite{ABN}) and Eq.~(\ref{eq:hypersq}):
\be
\frac{M_{B^*}^2-M_B^2}{M_{D^*}^2-M_D^2} = \frac{{\hat C}(m_b^2)}{{\hat C}(m_c^2)}
  \left[ 1 + \Lambda_{\rm eff} \left( \frac{1}{m_c}-\frac{1}{m_b} \right) + \ldots
    \right],
\label{ratrel} \ee
and $\Lambda_{\rm eff}$ in the subleading terms is a combination of the hadronic parameters. This mass splitting ratio is $0.8776$, based on the data \cite{PDG2025}. We now use in this relation the results (\ref{rathatC}), and we extract the value of this hadronic parameter
\begingroup\color{black}
\bea
\Lambda_{\rm eff} & = & \left[ 0.2358 \pm 0.0770(\A) ^{+0.0338}_{-0.0457}(M_1) ^{+0.0171}_{-0.0162}(\alpha_s) \pm 0.0006(\bam_b) \mp 0.0011(\bam_c) \right] \ {\rm GeV} = 0.236(91) \ {\rm GeV}.
\label{Leffres} \eea 
\endgroup

The quantities ${\hat C}(m_q^2)$ can formally be considered as observables because they are renormalisation scale (RScl) and scheme (RSch) invariant. The scale $\Lambda_{\rm eff}$, appearing in the relation (\ref{ratrel}), is thus also RScl and RSch invariant, thus formally an observable.

The perturbation series for the quantity $\cF_C(\sigma)$, Eq.~(\ref{FCTPSMSb}), in $\MSbar$ and where $\nu_0=1/3$ and $n_f=3$, turns out to have for $\sigma=m_b^2$ and $\sigma=m_c^2$ the smallest term $\sim a^{\nu_0+3}$ and $\sim a^{\nu_0+2}$, respectively, so we truncate the series at that term (including it). The corresponding uncertainty '(${\rm TPS}_{FCq}$)' is then taken if we add or subtract that (smallest) term to the mentioned truncated series. This, and using the TPS results (\ref{mcpT}) for the pole masses, leads to the following extracted values for $\cF_C(m_q^2)$ and the related hadronic quantities:
\bes
\label{FCsT}
\bea
\cF_C(m_b^2)_{\rm (TPS)} & = & 0.5058 \pm 0.0208 ({\rm TPS}_{FCb}) ^{-0.0034}_{+0.0037} ({\rm TPS}_{mb}) ^{+0.0041}_{-0.0040}(\alpha_s) \mp 0.0002 (\bam_b),
\label{FCmbT} \\
\cF_C(m_c^2)_{\rm (TPS)} & = & 0.6763 ^{+0.0920}_{-0.0919} ({\rm TPS}_{FCc}) ^{-0.0347}_{+0.0492} ({\rm TPS}_{mc}) ^{+0.0104}_{-0.0101}(\alpha_s) \mp 0.0011 (\bam_c).
\label{FCmcT}
\eea \ees
\bea
  \left(\frac{{\hat C}(m_b)}{{\hat C}(m_c)}\right)_{\rm (TPS)} = \left(\frac{ \cF_C(m_b^2) }{\cF_C(m_c^2)}\right)_{\rm (TPS)}
  &=& 0.7479 ^{-0.0895}_{+0.1176}({\rm TPS}_{FCc}) ^{+0.0307}_{-0.0308}  ({\rm TPS}_{FCb}) ^{+0.0404}_{-0.0507} ({\rm TPS}_{mc}) ^{-0.0051}_{+0.0054} ({\rm TPS}_{mb})
\nonumber\\ &&
  \mp 0.0054 (\alpha_s)  \mp 0.0003 (\bam_b) \pm 0.0011 (\bam_c).
\label{rathatCT} \eea
\bea
(\Lambda_{\rm eff})_{\rm (TPS)} & = & {\big [} 0.4024 \pm 0.3700 ({\rm TPS}_{FCc}) ^{-0.1075}_{+0.1168} ({\rm TPS}_{FCb}) ^{-0.0877}_{+0.0922}  ({\rm TPS}_{mc}) ^{+0.0129}_{-0.0138}  ({\rm TPS}_{mb})
\nonumber\\ &&
  ^{+0.0238}_{-0.0229} (\alpha_s) \pm 0.0006 (\bam_b) \mp 0.0022 (\bam_c) {\big ]} \ {\rm GeV}.
\label{LeffT} \eea 

In all these quantities, '(${\rm TPS}_{FCq}$)' denotes the uncertainty originating from the uncertainty TPS in the quantity $\cF_C(m_q^2)$, cf.~Eqs.~(\ref{FCsT}). Analogously, '(${\rm TPS}_{mq}$)' denotes the uncertainty originating from the uncertainty TPS in the quantity $m_q$, cf.~Eqs.~(\ref{mqpT}). All these uncertainties approximately reflect the corresponding leading IR-renormalon ambiguities of the quantities $\cF_C(m_q^2)$ and $m_q$. In this TPS approach, these ambiguities cannot be made correlated, for lack of more detailed knowledge.

In the above TPS results, Eqs.~(\ref{FCsT})-(\ref{LeffT}), we see that the mentioned uncertainties originating from the IR sector, '(${\rm TPS}_{FCq}$)' and '(${\rm TPS}_{mq}$)', are in general considerably larger than the IR-sector ambiguities '($M_1$)' in the renormalon-motivated resummation approach, cf.~Eqs.~(\ref{FCs})-(\ref{rathatC}), (\ref{Leffres}). One reason for this is that in the resummed approach the IR-sector uncertainties are all correllated to the common 3$\delta$AQCD running coupling (that is free of Landau singularities), i.e., we have these (expected) correlations under control by the common IR-regulated running coupling.

\subsubsection{Determination of ${\hat \mu}_G^2$ from hyperfine splitting.}

Using the measured values $M_{B^*}$, $M_B$, and the Wilson coefficient ${\hat C}(m_b^2)$, we determine from Eq.~(\ref{eq:hypersq}) the parameter ${\hat \mu}_G^2$
\begin{equation}
  \hat{\mu}_G^2 \simeq \frac{3}{4}\frac{M_{B^*}^2-M_B^2}{\hat{C}(m_b) \left(1-\Lambda_{\rm eff}\frac{1}{m_b} \right)} \approx \frac{3}{4}\frac{M_{B^*}^2-M_B^2}{\hat{C}(m_b)}\left(1+\Lambda_{\rm eff}\frac{1}{m_b} \right).
\label{hatmu2}  
\end{equation}
%The DSRS reference gives an example numerical determination $\mu_G^2 \approx 0.284 \pm 0.014~{\rm GeV}^2$ using ${\hat C}$ at N$^3$LO \cite{Hayashi:2023axn}.
We obtain, using the above values for $\Lambda_{\rm eff}$ and ${\hat{C}}(m_b^2)$
\begingroup\color{black}
\bea
  \hat{\mu}_G^2 & \approx &
  \frac{3}{4}\frac{M_{B^*}^2-M_B^2}{ {\hat{C}}(m_b^2)} \left(1+\Lambda_{\rm eff}\frac{1}{m_b} \right) 
 \nonumber\\
 & = & \left[ 0.5073 \mp 0.0066(\A) ^{-0.0046}_{+0.0039}(M_1) \mp 0.0035(\alpha_s) \pm 0.0002(\bam_b) \mp 0.0001(\bam_c) \right] \ {\rm GeV}^2 = 0.507(9)  \ {\rm GeV}^2.
\label{hatmu2mb}
 \eea
\endgroup
 We can also obtain the value of $\hat{\mu}_G^2$ from the charm system, using our values of $\Lambda_{\rm eff}$ and ${\hat{C}}(m_c^2)$
 \begingroup\color{black}
 \bea
 \hat{\mu}_G^2 &\approx& \frac{3}{4}\frac{M_{D^*}^2-M_D^2}{\hat{C}(m_c^2)}\left(1+\Lambda_{\rm eff}\frac{1}{m_c} \right) 
 \nonumber\\
 & = & \left[ 0.5053 \mp 0.0110(\A) ^{-0.0048}_{+0.0044}(M_1) \mp 0.0037(\alpha_s) \pm 0.0002(\bam_b) \mp 0.0001(\bam_c) \right] \ {\rm GeV}^2 = 0.505(13)  \ {\rm GeV}^2.
\label{hatmu2mc}
\eea
\endgroup
Comparing the two values, Eqs.~(\ref{hatmu2mb}) and (\ref{hatmu2mc}), we see that they are very close to each other. This represents a cross check of our numerical analysis, because the hadronic quantity $\hat{\mu}_G^2$ is scale (and scheme) invariant and heavy-quark symmetry universal. However, we trust more the value (\ref{hatmu2mb}), because the missing theoretical terms in Eq.~(\ref{hatmu2mb}) [of relative size $\sim (\Lambda_{\rm eff}/m_b)^2$] are very suppressed and much smaller than in Eq.~(\ref{hatmu2mc}) [of relative size $\sim (\Lambda_{\rm eff}/m_c)^2$].

\begingroup\color{black}
One may also be interested in the full (RScl-invariant) quantity $C(m_b^2;\mu^2) \mu_G^2(\mu^2) = {\hat C}(m_b^2) {\hat \mu}_G^2$ appearing in the heavy-quark expansion (\ref{eq:MHexp}) for $M_B$ and $M_{B^*}$. According to Eq.~(\ref{hatmu2mb}), it has $1/m_b$ expansion whose numerical evaluation gives
\bes
\label{Cmu2}
\bea
{\hat C}(m_b^2) {\hat \mu}_G^2 &=& \frac{3}{4}(M_{B^*}^2 - M_B^2)\left[1 + \frac{\Lambda_{\rm eff}}{m_b} + \cO ( ( \Lambda_{\rm eff}/m_b)^2 ) \right]
\label{Cmu2a} \\
&=& [ 0.362 + 0.017(6) + \ldots ] \ {\rm GeV}^2,
\label{Cmu2b} \eea \ees
where in the last line we used the results Eq.~(\ref{Leffres}) for $\Lambda_{\rm eff}$ and Eq.~(\ref{mbp}) for $m_b$. For the $D$-meson systems the corresponding numerical result is $[0.413+0.058(9)+ \ldots]$ ${\rm GeV}^2$.
\endgroup
  
\subsubsection{Determination of $\bar\Lambda$ and $\mu_\pi^2$ from spin-averaged masses.}

For this purpose we will use both spin-averaged experimental values of $\overline M_B$ and $\overline M_D$ in the general form, cf.~Eq.~(\ref{eq:spinave})
\begin{equation}
     \overline M_B = m_b + \bar\Lambda + \frac{\mu_\pi^2}{2m_b}
  + \mathcal{O}\!\left(\frac{\Lambda^3}{m_b^2}\right), \qquad \overline M_D = m_c + \bar\Lambda + \frac{\mu_\pi^2}{2m_c}
  + \mathcal{O}\!\left(\frac{\Lambda^3}{m_c^2}\right).
\label{bMH}
\end{equation}

Solving this system of equations at $\mathcal{O}(1/m_h)$ gives
\bes
\label{mupi2bL}
\bea
     \mu_\pi^2 &=&
  2 m_b m_c\left(1-\frac{\overline M_B-\overline M_D}
  {m_b-m_c}\right),
\label{mupi2} \\  
  \bar\Lambda &=& \overline M_B - m_b - \frac{\mu_\pi^2}{2m_b}
  = \overline M_D - m_c - \frac{\mu_\pi^2}{2m_c}.
\label{bL} \eea \ees  
Using our results for $m_b$ and $m_c$, Eqs.~(\ref{mqp}), we obtain
 \begingroup\color{black}
\bes
\label{mupi2bLnum}
\bea
\mu_\pi^2 &=& \left[ -0.0219 \pm 0.0591(\A) ^{-0.5684}_{+0.2620}(M_1) ^{+0.0706}_{-0.0592}(\alpha_s) ^{+0.0401}_{-0.0351}(\bam_b) ^{-0.0303}_{+0.0350}(\bam_c) \right] \ {\rm GeV}^2 = -0.022(577) \ {\rm GeV}^2,
\label{mupi2num} \\
\bar\Lambda &=&  \left[ 0.3080 \mp 0.0941(\A) ^{-0.0009}_{+0.0246}(M_1) ^{-0.0350}_{+0.0329}(\alpha_s) ^{-0.0120}_{+0.0105}(\bam_b) ^{+0.0030}_{-0.0035}(\bam_c) \right]  \ {\rm GeV} = 0.308(104) \ {\rm GeV}.
\label{bLnum} \eea \ees
\endgroup
As we can see, the result for the hadronic parameter $\mu_\pi^2$ is very unstable under variations of parameters ($M_1$, $\alpha_s$, $\bam_q$) because of the near perfect cancellation of the terms $+1$ and $-(\overline M_B-\overline M_D)/(m_b-m_c)$, and our results indicate that the value of $\mu_\pi^2$ is compatible with zero.

It is interesting that the numerical variations for the hadronic quantities ${\hat \mu}^2_G$ and ${\bar\Lambda}$ under the variation of the IR ambiguity (IR threshold) parameter $M_1=(0.150^{+0.100}_{-0.050})$ GeV are comparable to, or at least not significantly larger, that those coming from the variation of the (pQCD) parameter $\alpha_s^{\MSbar}(M_Z^2)=0.1180 \pm 0.0009$. This is not the case for the quark pole masses $m_b$ and $m_c$ where the IR-ambiguity variations [the ('$M_1$') variations] are clearly the dominant ones, cf.~Eqs.~(\ref{mqp}), especially in the case of $m_c$ mass. 

For completeness, we also give here the values of the quantities $\hat{\mu}_G$, $\mu^2_{\pi}$ and $\bar\Lambda$ as obtained from the quantites $m_q$, $\hat C(m_q)$ and $\Lambda_{\rm eff}$ extracted from the TPS approach [Eqs.~(\ref{mqpT}) and (\ref{FCsT})-(\ref{LeffT})]
\bea
  (\hat{\mu}_G^2)_{\rm (TPS)} & \approx &
  \frac{3}{4}\frac{M_{B^*}^2-M_B^2}{ {\hat{C}}(m_b)} \left(1+\Lambda_{\rm eff}\frac{1}{m_b} \right) 
 \nonumber\\
 & = &
     {\big [} 0.5282 ^{+0.0362}_{-0.0361} ({\rm TPS}_{FCc}) ^{-0.0105}_{+0.0115} ({\rm TPS}_{FCb}) ^{-0.0085}_{+0.0092}  ({\rm TPS}_{mc})  ^{+0.0037}_{-0.0039}({\rm TPS}_{mb})
 \nonumber\\ &&
       ^{-0.0020}_{+0.0022} (\alpha_s)  \pm 0.0002 (\bam_b) \mp 0.0002 (\bam_c)  {\big ]} \ {\rm GeV}^2.
\label{hatmu2mbT}
 \eea
 \bea
 (\hat{\mu}_G^2)_{\rm (TPS)} &\approx& \frac{3}{4}\frac{M_{D^*}^2-M_D^2}{\hat{C}(m_c)}\left(1+\Lambda_{\rm eff}\frac{1}{m_c} \right) 
 \nonumber\\
 & = &
     {\big [} 0.5224 ^{+0.0973}_{-0.0972} ({\rm TPS}_{FCc}) ^{-0.0282}_{+0.0307} ({\rm TPS}_{FCb}) ^{-0.0058}_{+0.0049}  ({\rm TPS}_{mc})  ^{+0.0034}_{-0.0036}({\rm TPS}_{mb})
\nonumber\\ &&
       ^{-0.0026}_{+0.0027} (\alpha_s) \pm 0.0002 (\bam_b) \mp 0.0001 (\bam_c)  {\big ]} \ {\rm GeV}^2.
\label{hatmu2mcT}
\eea
\bes
\label{mupi2bLnumT}
\bea
(\mu_\pi^2)_{\rm (TPS)} &=& \left[ +0.3555 ^{-1.0525}_{+0.7096} ({\rm TPS}_{mc}) ^{+0.7134}_{-0.7341}  ({\rm TPS}_{mb}) ^{+0.0504}_{-0.0491} (\alpha_s) \pm 0.0323 (\bam_b) \mp 0.0216 (\bam_c) \right]  \ {\rm GeV}^2,
\label{mupi2numT} \\
(\bar\Lambda)_{\rm (TPS)} &=&  \left[ 0.2743 ^{+0.1051}_{-0.0709} ({\rm TPS}_{mc}) ^{-0.2250}_{+0.2316} ({\rm TPS}_{mb}) ^{-0.0279}_{+0.0268} (\alpha_s) \mp 0.0102(\bam_b) \pm 0.0022 (\bam_c) \right] \ {\rm GeV}. 
\label{bLnumT} \eea \ees
Again, when we compare, in these TPS-extracted values, the uncertainties originating from the IR sector (IR-renormalon ambiguities), '(${\rm TPS}_{FCq}$)' and  '(${\rm TPS}_{mq}$)', we see that they are considerably larger than the IR-sector ambiguities '($M_1$)' in the renormalon-motivated resummation approach, cf.~Eqs.~(\ref{hatmu2mb})-(\ref{hatmu2mc}) and (\ref{mupi2bLnum}).

\section{Comparisons and Conclusions}
\label{sec:concl}

We applied a renormalon-motivated resummation approach, together with a QCD running coupling regularised in the IR (free of Landau singularities), to the timelike hadronic quantities such as the quark pole mass ($m_b$ and $m_c$), the chromomagnetic Wilson coefficient (${\hat C}(m_c^2)$ and ${\hat C}(m_b^2)$), and several associated hadronic quantities ($\Lambda_{\rm eff}$, ${\hat{\mu}}_G^2$, $\bar {\Lambda}$). The IR ambiguities in our approach were parametrised with the variation of the  IR-threshold parameter values $M_1 \sim m_{\pi}$ ($M_1=0.150^{+0.100}_{-0.050}$ GeV) of the lattice-motivated 3$\delta$AQCD QCD coupling (free of Landau singularities). The obtained numerical results are given in Eqs.~(\ref{mqp}), (\ref{FCs})-(\ref{Leffres}), (\ref{hatmu2mb})-(\ref{hatmu2mc}) and (\ref{mupi2bLnum}). While these extracted numerical results show in general considerable uncertainty due to the mentioned IR-regime uncertainty (variation of $M_1$), the extracted values of the (measurable) hadronic parameters ${\hat{\mu}}_G^2$, as well as of ${\bar \Lambda}$, show the IR-regime uncertainty not significantly larger than the uncertainty originating from the (world average) variation of the pQCD input parameter $\alpha_s^{\MSbar}(M_Z^2)=0.1180 \pm 0.0009$.

We can compare these results with those of another approach \cite{AyaPin2019,AyaPin2026}, that uses hyperasymptotic expansion for the Principal Value in pQCD \cite{AyaPin2019b}. Their obtained numerical results, while different, are in general compatible with our results. For example, our result for the ratio of the chromomagnetic Wilson coefficients is \textcolor{black}{${\hat C}(m_b^2)/{\hat C}(m_c^2) =0.802(24)$ (and $\Lambda_{\rm eff}=0.236(91)$ GeV)}, and theirs is $0.836(32)$ (and $\Lambda_{\rm eff}=-A=0.121(85)$ GeV); our result for ${\hat{\mu}}_G^2$ obtained from the $B$ system Eq.~(\ref{hatmu2mb}) is \textcolor{black}{${\hat{\mu}}_G^2=0.507(9) \ {\rm GeV}^2$,} and theirs is $0.507(7)$. The two results for the (measurable) quantity ${\hat{\mu}}_G^2$ show remarkable consistency.
We point out, however, that the obtained values of the quark pole masses in \cite{AyaPin2019,AyaPin2026} are clearly different from ours [Eqs.~(\ref{mqp})], \textcolor{black}{but still marginally compatible: $m_{b,\rm PV}=4.843(37)$ GeV (ours: $m_b=5.008(110)$ GeV) and $m_{c,\rm PV}=1.427(42)$ GeV (ours: $m_c=1.670(189)$ GeV).}

Furthermore, since our approach, unlike that of \cite{AyaPin2026}, is independent of the renormalisation scale (RScl), as pointed out in the paper, among the uncertainties of our extracted values there is none that would have its origin in the variation of RScl. \textcolor{black}{Further, since we use AQCD-couplings (that are free of Landau singularities) in our resummation, we have formally no (IR) renormalon ambiguities. However, analogs of such RScl- and renormalon ambiguities may be indirectly reflected, in our approach, in the details of the (spacelike and timelike) AQCD coupling behaviour in the IR regime. In particular, they may be reflected in the uncertainties '$(M_1)$' obtained by varying the IR threshold parameter $M_1$ ($\sim m_{\pi}$) of the 3$\delta$AQCD coupling, and in the uncertainties '$(\A)$' obtained by going from 3$\delta$AQCD to 2$\delta$AQCD couplings as described in the text. Furthermore, part of the uncertainties analogous to the renormalon ambiguities are reflected in} the uncertainties '(an)' coming from the nonuniqueness of our ans\"atze for the modified Borel transforms $\tB[\cF_q](u)$ ($\Rightarrow \tB[\cD_q](u)$) for $m_q$, Eqs.~(\ref{tBFq})-(\ref{tBDans}), and of the modified modified Borel transforms $\tB[\cF_C](u)$ ($\Rightarrow \tB[\cD_C^{(1)}](u)$) [Eqs.~(89)-(95) of Ref.~\cite{JPG2026}] for ${\hat C}(m_q^2)$. \textcolor{black}{For $m_q$, such '(an)' uncertainties are relatively small. Since we do not know the details of the renormalon structure beyond the $u=1/2$ IR renormalons of the transforms $\tB[\cF_C](u)$, we have not estimated the '(an)' uncertainties for the quantities ${\hat C}(m_q^2)$ and the related quantities.}

In our approach, we regarded the uncertainties from the IR-regime '($M_1$)', as well as those from the pQCD coupling value '($\alpha_s$)', to be correlated in the quark pole masses $m_q$ and the mentioned hadronic quantites (${\hat C}(m_q^2)$, ${\hat{\mu}}_G^2$, etc.), because the values of the parameters $M_1$ and $\alpha_s^{\MSbar}(M_Z^2)$ are the two input values that determined our QCD running coupling (free of Landau singularities) that entered the evaluations of all these hadronic quantities. \textcolor{black}{Further, the uncertainties '$(\A)$', reflecting the general uncertainties of the coupling in the IR regime, were calculated in the hadronic quantities ${\hat C}(m_q^2)$ (and the related quantities) as the difference between the central case of the 3$\delta$AQCD coupling and the case of a 2$\delta$AQCD coupling, and again, the values of the masses of $m_q$ that are used in these quantities in 2$\delta$AQCD are those obtained in 2$\delta$AQCD.}

\textcolor{black}{Our analysis of the '$(\A)$' uncertainties suggests that the 3$\delta$AQCD frameworks, which follow closely the behaviour of a product of the (Landau-gauge) gluon and ghost dressing functions as determined by large-volume lattice calculations in the literature [cf.~\cite{3dLVLattice}], give us a realistic description of the Landau-ghost-free couplings in the IR regime, when we choose the values of the IR threshold mass $M_1$ of the spectral function of the coupling in the natural range $M_1 \sim m_{\pi}$.}

We note that the charm quark decoupling in the quantities $m_b$ and ${\hat C}(m_b^2)$ (i.e., the use of $n_f=3$ in them, instead of $n_f=4$) does influence our numerical analysis somewhat.\footnote{If excluding the '$(\A)$' uncertainties, we obtained  here ${\hat C}(m_b^2)/{\hat C}(m_c^2) =0.802(9)$ and $\Lambda_{\rm eff}=0.236(49)$ GeV [Eqs.~(\ref{rathatC}) and (\ref{Leffres})]; while in \cite{JPG2026} where we used $n_f=4$ for ${\hat C}(m_b^2)$, we obtained ${\hat C}(m_b^2)/{\hat C}(m_c^2) =0.776(22)$ and $\Lambda_{\rm eff}=0.335(75)$ GeV [Eqs.~(107a) and (109a) there].}

The results change significantly when truncated perturbation series (TPS) are used, for $\cF_q$ (i.e., $m_q$) as well as for $\cF_C$ (i.e., ${\hat C}$),
cf.~Eqs.~(\ref{mqpT}), (\ref{FCsT})-(\ref{LeffT}), (\ref{hatmu2mbT})-(\ref{mupi2bLnumT});
the uncertainties originating in the truncation uncertainty\footnote{
As argued in the text, the truncation uncertainty in the TPS results reflects, indirectly, the leading IR renormalon ambiguity or, in our case, the IR-parameter '($M_1$)' ambiguity. It also includes, at least partly, the uncertainty due to RScl-variation.}
are much larger than the IR-regime uncertainties '($M_1$)' in our resummed approach. The quantities ${\hat C}(m_q^2)$ and $\Lambda_{\rm eff}$ were first evaluated in \cite{GandN}, with TPS approach (in $\MSbar$), with two terms in ${\hat C}(m_c^2)$ and three terms in ${\hat C}(m_b^2)$ (i.e., truncated at $\sim a^{\nu_0+1}$ and $\sim a^{\nu_0+2}$, respectively), and using $n_f=4$ in both cases. Furthermore, they used the values $\alpha_s^{\MSbar}(m_c^2)=0.36$ and $\alpha_s^{\MSbar}(m_b^2)=0.22$, and obtained for the ratio the value ${\hat C}(m_b)/{\hat C}(m_c) \approx 0.697/0.847 = 0.823$ [cf.~their Table 1 and Eq.~(40) there], and consequently for the contribution $\Lambda_{\rm eff} (1/m_c-1/m_b)$ to Eq.~(\ref{ratrel}) the value of about 11 \% (we got about $(9.4 \pm 1.1)$ \% in our resummed approach). Uncertainties were not estimated there.

In addition, the parameter ${\hat{\mu}}_G^2$ was evaluated also in \cite{GMS,Betal,HST}, where the uncertainties are larger than here (in the resummed approach) and \cite{AyaPin2026}, primarily due to the truncation effects.

\acknowledgments
This work was supported in part by FONDECYT (Chile) Grants No.~1220095 (G.C.) and 1240329 (C.A.).

%\appendix
%\section{}

% Bibliography placeholders (replace with your own BibTeX entries / arXiv refs)

\end{document}